\documentclass{ws-procs9x6}
\def\trimmarks{}
\usepackage{cite}

\ifx\ensuremath\undefined%
  \def\ensuremath#1{\ifmath{#1}}
\fi

\def\W{\ensuremath{\mathrm{W}}}

\def\Hpm{\ensuremath{\mathrm{H}^\pm}}
\def\HpHm{\ensuremath{\mathrm{H}^+\mathrm{H}^-}}

\def\Hpmpm{\ensuremath{\mathrm{H}^{\pm\pm}}}

\def\Zo{\ensuremath{\mathrm {Z}}}
\def\Ho{\ensuremath{\mathrm {H}}}
\def\GeV{\ensuremath{\mathrm {GeV}}}

\def\ra{\ensuremath{\rightarrow}}

\def\epem{\ensuremath{\mathrm{e}^+\mathrm{e}^-}}

\begin{document}
\title{Searches for Exotic Higgs Bosons at LEP
\footnote{\uppercase{T}alk presented at {\it \uppercase{SUSY} 2003: \uppercase{S}upersymmetry in the \uppercase{D}esert}\/, held at the \uppercase{U}niversity of \uppercase{A}rizona, \uppercase{T}ucson, \uppercase{AZ}, \uppercase{J}une 5-10, 2003. \uppercase{T}o appear in the \uppercase{P}roceedings.}
}

%----------------------------------------------------------------------

\author{ANDRE GEORG HOLZNER}

\address{
  Institut f\"ur Teilchenphysik,
  ETH H\"onggerberg\\
  CH-8093 Z\"urich, Switzerland\\
  E-mail: Andre.Holzner@cern.ch\\
    }

%----------------------------------------------------------------------

\maketitle

\abstracts{
In the Standard Model (SM), the weak gauge bosons and fermions acquire mass through the 
Higgs mechanism. A lower limit on the SM Higgs mass of 114.4~\GeV{}
was obtained from the direct search at LEP. Although a single Higgs doublet is sufficient to explain
the non-zero particle masses while keeping the theory SU(2) gauge invariant, several extensions to this
minimal model were proposed, to which this limit does not apply. Most of the models discussed here introduce one
additional Higgs doublet and are therefore called {\sl Two Higgs doublet models} (2HDM). 
Several signatures predicted by such models have been searched for at LEP
using
data collected at center-of-mass energies up to 209~\GeV{}.
All limits quoted in this report are at 95\% confidence level.
}

%----------------------------------------------------------------------
\section{Fermiophobic Higgs bosons}

In some scenarios, the Higgs to fermion coupling can
vanish at tree level~\cite{Brucher:1999tx}. 
At low masses, such a Higgs particle decays predominantly into photons
while at higher masses (when it becomes kinematically possible), it
decays preferably into a pair of \W{} or \Zo{} bosons (where one of
the bosons can be off-shell). 
All searches for bosonic Higgs decays reported here assume that the 
production mechanism is \epem \ra{} \Ho\Zo.

%--------------------

\subsection{Higgs decays into photons}

\begin{figure}[ht]
\centerline{
\epsfxsize=4cm\epsfbox{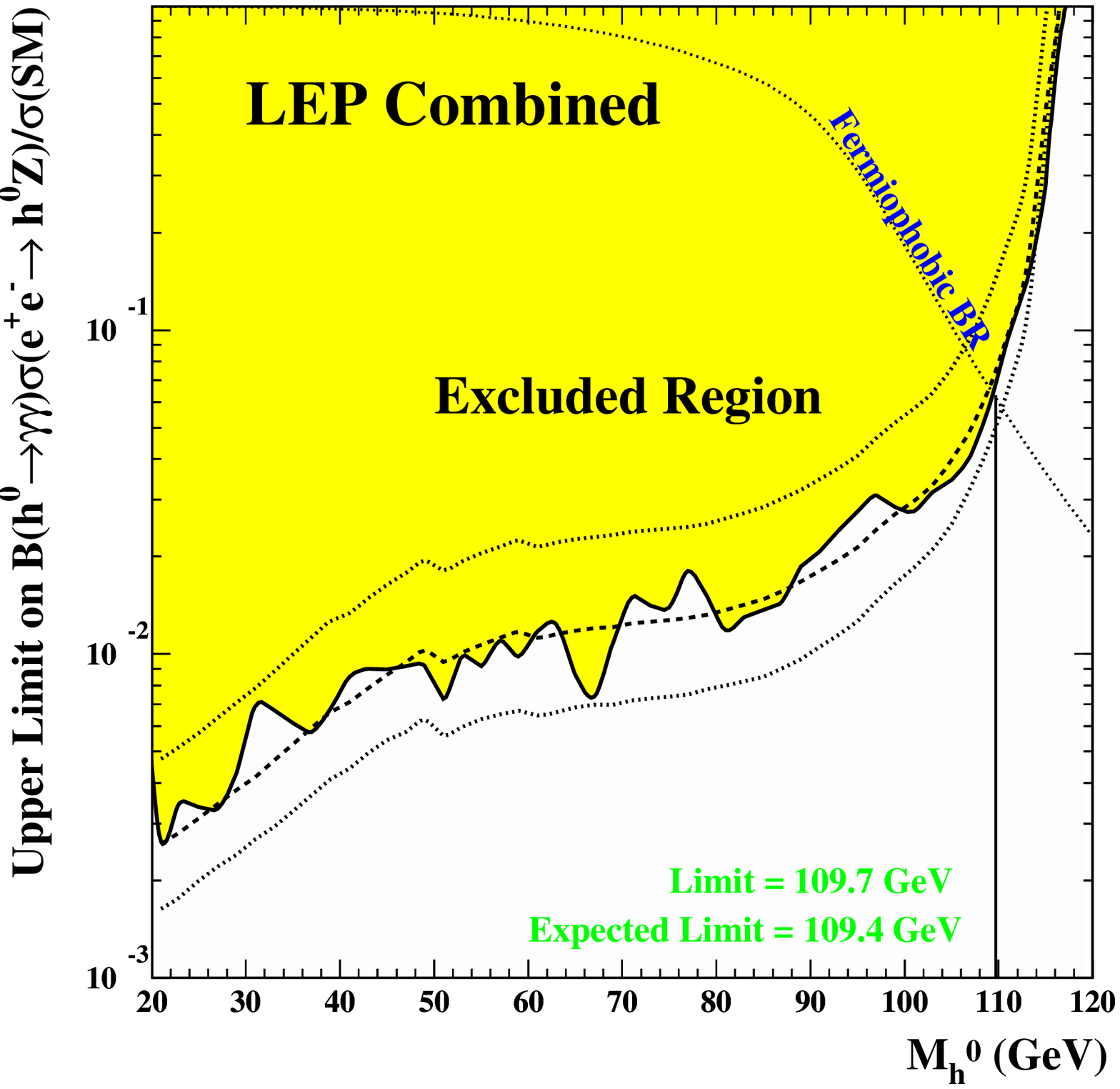}
\epsfxsize=6cm\epsfbox{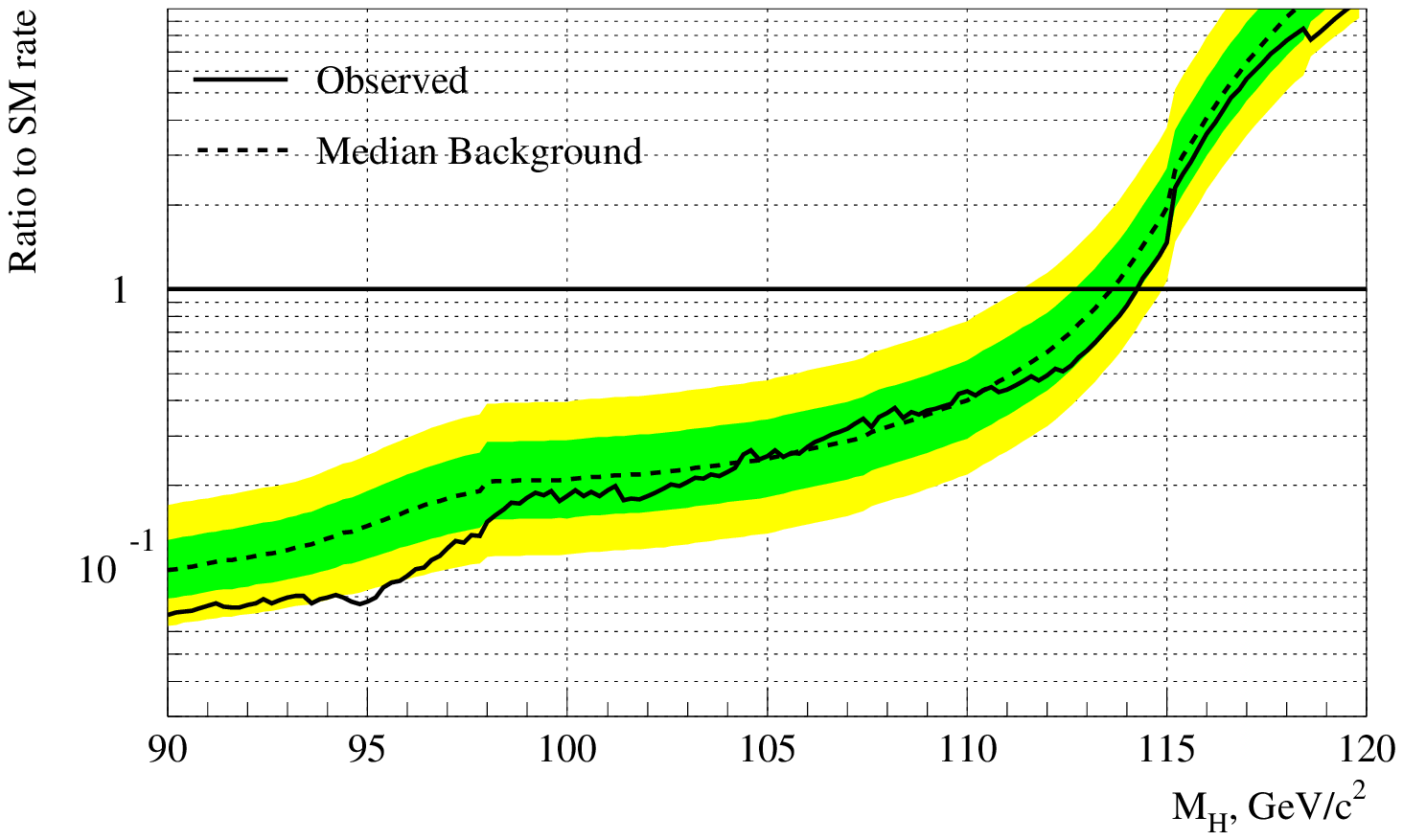}
}
\caption{\label{fig:lep-phobic-and-invisible}
Left: Excluded regions of the Higgs mass versus
the production cross section times branching ratio into photons
normalized to the SM Higgs production cross
section for the combination of the four LEP experiments' Higgs to
photon decay searches~\cite{lep-phobic-2002}. Right: The excluded Higgs
production cross section (divided by the SM Higgs-Strahlung cross
section) as function of the Higgs mass hypothesis for invisibly
decaying Higgs bosons (LEP combination)~\cite{:2001xz}.}

\end{figure}

Combining the two photons from the Higgs decay with the \Zo{} decay
modes, one obtains three different final states:
$\gamma\gamma\mathrm{qq}$,
$\gamma\gamma\nu\nu$ and
$\gamma\gamma\ell\ell$.
All four LEP collaborations have eagerly searched
for the corresponding signatures but no significant deviation from the SM
backgrounds was observed. 
This non-observation is 
quantified as excluded region in the plane of the Higgs mass versus
the production cross section times branching ratio into photons
normalized to the SM Higgs production cross
section~\cite{lep-phobic-2002}, as shown in
Fig.~\ref{fig:lep-phobic-and-invisible} (left). A 
(preliminary)  
lower limit is set on the Higgs mass at 109.7
\GeV{} within the fermiophobic benchmark scenario~\cite{lep-phobic-2002}.

%--------------------

\subsection{Higgs decays into \W{} and \Zo{} bosons}

Because three gauge bosons are produced in this channel ($\W\W^*\Zo$ or
$\Zo\Zo^*\Zo$), one gets a variety of 
6-fermion final states. Six categories of these final states have been
analyzed by the L3 collaboration, covering 93\% of the total branching ratio. Due to the absence
of any indication for the existence of such a  Higgs, a lower limit on its
mass at 107~\GeV{} is set~\cite{Achard:2003jb}.

%----------------------------------------------------------------------
\section{Invisibly decaying Higgs bosons}

Here we consider the possibility that Higgs boson decays fully or
partially to invisible particles. As an example, some specific sets of the parameters of the Minimal Supersymmetric
Standard Model (MSSM) predict Higgs decays into neutralinos. 
Again, searches at LEP assume that such a
Higgs is produced in association with a \Zo{} boson, leading to the 
two experimental signatures ``two jets plus missing energy'' and ``two
charged leptons plus missing energy''. In neither of these two
channels was any evidence for Higgs production observed~\cite{:2001xz}. The excluded Higgs
production cross section (divided by the SM Higgs-Strahlung cross
section) as function of the Higgs mass hypothesis is shown in
Fig.~\ref{fig:lep-phobic-and-invisible} (right). In models for which the Higgs production
cross section is equal to the SM Higgs production cross section, Higgs
masses below 114.4 \GeV{} (preliminary) are excluded.

%----------------------------------------------------------------------
\section{Charged Higgs bosons \Hpm}

In the MSSM, the \Hpm{} bosons are heavier than the \W{}
(neglecting radiative corrections).  
Strict limits arise from $\mathrm{b} \ra \mathrm{s}\gamma$
measurements~\cite{Ciuchini:1997xe},
which, however, do  not apply to so-called type I 2HDMs.
The process \epem \ra{} \HpHm\  becomes kinematically possible at
LEP center-of-mass energies.
The experimental signatures depend strongly on
the value of $\tan\beta$ (the ratio of the vacuum expectation values of the two
Higgs doublets): For small values of $\tan\beta$, the main \Hpm{} decay modes 
are into $\mathrm{cs}$ and $\tau\nu_\tau$
while for large $\tan\beta$ so-called `three body decays' 
($\mathrm{H} \ra \mathrm{A}\mathrm{q}\bar{\mathrm{q}},
\mathrm{A}\ell\nu$) via a virtual \W{} become dominant.

%--------------------
  
\subsection{Small values of $\tan\beta$}

All four LEP collaborations have searched for pair production of
charged Higgs bosons, in the final states 
$\mathrm{c}\mathrm{s}\mathrm{c}\mathrm{s}$, 
$\mathrm{c}\mathrm{s}\tau\nu$ and
$\tau\nu\tau\nu$. No significant excess was observed. 

A preliminary lower limit on the Higgs mass is 78.6~\GeV{},
independent of the branching ratio into $\tau\nu$\cite{lep-charged-2001}.

%----------------------------------------
\subsection{Large values of $\tan\beta$}

Due to the nature of the three body decays of the charged Higgs
bosons at high values of $\tan\beta$, the possible experimental
signatures become more complicated. 
For example, OPAL performs a detailed analysis explicitly looking for 
the final states 
$(\mathrm{qqbb})(\mathrm{qqbb})$,
$(\ell\nu\mathrm{bb})(\mathrm{qqbb})$ and 
$(\tau\nu)(\mathrm{qqbb})$. 
No indication of a signal has been found. The lower limit of these two analyses is 
at 76.6~\GeV{} (preliminary)~\cite{delphi-wawa,opal-wawa}.

%----------------------------------------------------------------------
\section{Doubly charged Higgs bosons \Hpmpm}

Such Higgs particles can arise e.g.\ in Higgs triplet or left-right
symmetric models.
Charge conservation demands that doubly charged Higgs bosons only couple to
charged leptons and other gauge and Higgs bosons at tree level. 

One has to distinguish between three different experimental topologies 
which correspond to different ranges of the Higgs life time: Very
short lifetimes lead to a four-lepton signature, intermediate
lifetimes result in a Higgs decaying within the tracking chambers
(i.e.\  tracks with kinks) while for longer lifetimes, two heavy
doubly charged particles are seen in the detector. Three LEP
experiments have searches for pair production of doubly charged Higgs 
bosons~\cite{Abdallah:2002qj,Achard:2003mv,Abbiendi:2001cr},
without observing any evidence for such a signal. The lowest
limit of these searches is set at 95.5~\GeV{}.

%----------------------------------------------------------------------
\section{Summary}

Various models involving an extended Higgs sector have been probed
at LEP. None of the searches has shown some evidence for such an
extension of the SM. Limits have been set at 95\% confidence level, by each
collaboration and --- where possible --- for the combined LEP
data. Most of the results are still preliminary, the LEP combination
of final results will hopefully follow soon.

%----------------------------------------------------------------------

\end{document}